\documentclass[12pt]{article}

\usepackage{amsfonts, amsmath}

\newcommand{\be}{\begin{equation}} \newcommand{\ee}{\end{equation}}

\begin{document}
\title{Non-Unitary and Unitary Transitions
 in Generalized Quantum  Mechanics and Information Problem Solving}
\thispagestyle{empty}

\author{A.E.Shalyt-Margolin\hspace{1.5mm}\thanks
{Fax: (+375) 172 326075; e-mail: a.shalyt@mail.ru;alexm@hep.by}}
\date{}
\maketitle
 \vspace{-25pt}
{\footnotesize\noindent  National Center of Particles and High
Energy Physics, Bogdanovich Str. 153, Minsk 220040, Belarus\\
{\ttfamily{\footnotesize
}}

\rm\normalsize \vspace{0.5cm}
\begin{abstract}
The present work is a study of the unitarity problem for Quantum
Mechanics at Planck Scale considered as Quantum Mechanics with
Fundamental Length (QMFL).In the process QMFL is described as
deformation of a well-known Quantum Mechanics (QM).\\Similar to
previous works of the author, the basic approach is based on
deformation of the density matrix (density pro-matrix) with
concurrent development of the wave function deformation in the
respective Schrodinger picture. It is demonstrated that the
existence of black holes in the suggested approach in the end
twice results in nonunitary transitions (first after the Big Bang
of QMFL to QM, and then when on trapping of the matter into the
black hole the situation is just the opposite - from QM to
QMFL)and hence in recovery of the unitarity. In parallel this
problem is considered in the deformation terms of Heisenberg
algebra, showing the  identity of the basic results. From this an
explicit solution for Hawking's informaion paradox has been
derived.
\end{abstract}
\newpage
\section{Introduction}
As is known, the Early Universe Quantum Mechanics
(Quantum Mechanics at Planck scale) is distinguished from a well-
known Quantum Mechanics at conventional scales
\cite{r1},\cite{r2}by the fact that in the first one the
Generalized Uncertainty Relations (GUR)are fulfilled resulting in
the emergence of a fundamental length, whereas in the second one
the usual Heisenberg Uncertainty Relations are the case.
In case of Quantum Mechanics with Fundamental
Length (QMFL) all three well-known fundamental constants
are present $G$,$c$ è $\hbar$, while the classical QM is associated
only with a single one $\hbar$. It is obvious that
transition from the first to the second one within the inflation
expansion is a nonunitary process, i.e. the process where the
probabilities are not retained
\cite{r3}, \cite{r4}.

Because of this, QMFL is considered as a deformation of QM. The
deformation in Quantum Mechanics at Planck scale takes different
paths: commutator deformation or more precisely deformation of the
respective Heisenberg algebra \cite{r5},\cite{r6},\cite{r7} , i.e.
the density matrix deformation approach,developed by the author
with co-workers in a number of papers
\cite{r3},\cite{r4},\cite{r8},\cite{r9},\cite{r10}. The first
approach suffers from two serious disadvantages: (1)the
deformation parameter is a dimensional variable
 $\kappa$ with a dimension of mass \cite{r5}; (2)in the limiting
transition to QM this parameter goes to infinity and fluctuations
of other values are hardly sensitive to it. Being devoid of the
above limitation, the second approach by the author's opinion is
intrinsic for QMFL: with it in QMFL the deformation parameter is
represented by the dimensionless quantity
$\alpha=l_{min}^{2}/x^{2}$, where $x$ is the scale and the
variation interval $\alpha$ is finite $0<\alpha\leq1/4$ \cite{r3},
\cite{r4},\cite{r10}. Besides, this approach contributes to the
solution of particular problems such as the information paradox
problem of black holes \cite{r3} and also the problem of an extra
term in Loiuwille equation \cite{r8},\cite{r9},\cite{r10},
derivation of Bekenstein-Hawking formula from the first principles
\cite{r10}, hypothesis of cosmic censorship \cite{r9},\cite{r10},
more exact definition and expansion of the entropy notion through
the introduction of the entropy density per unit minimum area
\cite{r9},\cite{r10},\cite{r11}. Moreover, it is demonstrated that
there exists a complete analogy in the construction and properties
of quantum mechanics and statistical density matrices at Planck
scale (density pro-matrices). It should be noted that an ordinary
statistical density matrix appears in the low-temperature limit
(at temperatures much lower than the Planck's)\cite{r12}.In the
present work the unitarity problem for QMFL is considered on the
basis of this approach. It is shown that as distinct from
Hawking's approach, in this treatment the existence of black holes
is not the  reason for the unitarity violation, rather being
responsible for its recovery. First after the Big Bang (initial
singularity)expansion of the Universe is associated with the
occurrence of a nonunitary transition from QMFL to QM, and with
trapping of the matter by the black hole (black hole
singularity)we have a reverse nonunitary process from QM to QMFL.
In such a manner a complete transition process from QMFL to the
unitarity may be recovered. Thus, the existence of black holes
contributes to the reconstruction of a symmetry of the general
picture. Similar results may be obtained in terms of the
Heisenberg's algebra deformation. So the problem of Hawking
information paradox is solved by the proposed approach: the
information quantity in the Universe is preserved. This paper is a
summing-up of the tentative results obtained by the author on the
information paradox as an extension of the earlier works \cite{r3}
è \cite{r11}.

\section {Some Preliminary Facts}

In this section the principal features of QMFL construction
are briefly outlined first in terms of the density matrix deformation
(von Neumann's picture) and subsequently in terms of the wave function deformation (Schrodinger picture) \cite{r3},\cite{r4},\cite{r9},\cite{r10}.
As mentioned above, for the fundamental deformation parameter
we use $\alpha = l_{min}^{2 }/x^{2 }$, where $x$ is the scale.
\\
\\
\noindent {\bf Definition 1.} {\bf(Quantum Mechanics with
Fundamental Length [for Neumann's picture])}
\\
\\
\noindent Any system in QMFL is described by a density pro-matrix
of the form $${\bf
\rho(\alpha)=\sum_{i}\omega_{i}(\alpha)|i><i|},$$ where
\begin{enumerate}
\item $0<\alpha\leq1/4$;
\item The vectors $|i>$ form a full orthonormal system;
\item $\omega_{i}(\alpha)\geq 0$ and for all $i$  the
finite limit $\lim\limits_{\alpha\rightarrow
0}\omega_{i}(\alpha)=\omega_{i}$ exists;
\item
$Sp[\rho(\alpha)]=\sum_{i}\omega_{i}(\alpha)<1$,
$\sum_{i}\omega_{i}=1$;
\item For every operator $B$ and any $\alpha$ there is a
mean operator $B$ depending on  $\alpha$:\\
$$<B>_{\alpha}=\sum_{i}\omega_{i}(\alpha)<i|B|i>.$$

\end{enumerate}
Finally, the following condition must be fulfilled:
\begin{equation}\label{U1}
Sp[\rho(\alpha)]-Sp^{2}[\rho(\alpha)]\approx\alpha.
\end{equation}
Consequently we can find the value for $Sp[\rho(\alpha)]$ satisfying the
condition of definition 1:
\begin{equation}\label{U2}
Sp[\rho(\alpha)]\approx\frac{1}{2}+\sqrt{\frac{1}{4}-\alpha}.
\end{equation}

According to point 5), $<1>_{\alpha}=Sp[\rho(\alpha)]$. Therefore
for any scalar quantity $f$ we have $<f>_{\alpha}=f
Sp[\rho(\alpha)]$. We denote the limit
$\lim\limits_{\alpha\rightarrow 0}\rho(\alpha)=\rho$ as the
density matrix. Evidently, in the limit $\alpha\rightarrow 0$ we
return to QM.

\renewcommand{\theenumi}{\Roman{enumi}}
\renewcommand{\labelenumi}{\theenumi.}

\renewcommand{\labelenumii}{\theenumii.}

As was shown in \cite{r3},\cite{r9},\cite{r10}:

\begin{enumerate}
\item The above limit covers both Quantum
and Classical Mechanics.
\item Density pro-matrix $\rho(\alpha)$ tests singularities.
As a matter of fact, the deformation parameter $\alpha$
should assume the value $0<\alpha\leq1$.  However, as seen from
(\ref{U2}), $Sp[\rho(\alpha)]$ is well defined only for
$0<\alpha\leq1/4$, i.e. for $x=il_{min}$ and $i\geq 2$ we have no
problems at all. At the point, where $x=l_{min}$ (that corresponds
to a singularity of space), $Sp[\rho(\alpha)]$ takes the complex
values.
\item It is possible to read the equation (\ref{U1}) as
\begin{equation}\label{U3}
Sp[\rho(\alpha)]-Sp^{2}[\rho(\alpha)]=\alpha+a_{0}\alpha^{2}
+a_{1}\alpha^{3}+...
\end{equation}
Then for example, one of the solutions of (\ref{U1}) is
\begin{equation}\label{U4}
\rho^{*}(\alpha)=\sum_{i}\alpha_{i} exp(-\alpha)|i><i|,
\end{equation},
where all
$\alpha_{i}>0$ are independent of $\alpha$  and their sum is equal
to 1 . In this way $Sp[\rho^{*}(\alpha)]=exp(-\alpha)$.
 Note that in the momentum representation $\alpha=p^{2}/p^{2}_{max}$,
 $p_{max}\sim p_{pl}$,where $p_{pl}$ is the Planck momentum.
 When present in matrix elements, $exp(-\alpha)$ can damp the
 contribution of great momenta in a perturbation theory. The solution
(\ref{U1})given by the formula (\ref{U4}) is further referred to as
{\bf(exponential ansatz)}.This ansatz will be the principal one in  our further
consideration.
\end{enumerate}
In\cite{r9},\cite{r10} it has been demonstrated, how a transition
from Neumann's picture to Shr{\"o}dinger's picture, i.e. from
the density matrix deformation to the wave function deformation,
may be realized by the proposed approach
\noindent {\bf Definition 2.} {\bf(Quantum Mechanics with
Fundamental Length [Shr{\"o}dinger's picture])}
\\
\\
Here, the prototype of Quantum Mechanical normed wave function (or
the pure state prototype) $\psi(q)$ with
$\int|\psi(q)|^{2}dq=1$ in QMFL is $\theta(\alpha)\psi(q)$. The
parameter of deformation $\alpha$ assumes the value
$0<\alpha\leq1/4$. Its properties are
$|\theta(\alpha)|^{2}<1$,$\lim\limits_{\alpha\rightarrow
0}|\theta(\alpha)|^{2}=1$ and the relation
$|\theta(\alpha)|^{2}-|\theta(\alpha)|^{4}\approx \alpha$ takes
place. In such a way the total probability always is less than 1:
$p(\alpha)=|\theta(\alpha)|^{2}=\int|\theta(\alpha)|^{2}|\psi(q)|^{2}dq<1$
and it tends to 1 when  $\alpha\rightarrow 0$. In the most general
case of the arbitrarily normed state in QMFL(mixed state prototype)
$\psi=\psi(\alpha,q)=\sum_{n}a_{n}\theta_{n}(\alpha)\psi_{n}(q)$
with $\sum_{n}|a_{n}|^{2}=1$ the total probability is
$p(\alpha)=\sum_{n}|a_{n}|^{2}|\theta_{n}(\alpha)|^{2}<1$ and
 $\lim\limits_{\alpha\rightarrow 0}p(\alpha)=1$.

 It is natural that Shrodinger equation is also
deformed in QMFL. It is replaced by the equation
\begin{equation}\label{U24}
\frac{\partial\psi(\alpha,q)}{\partial t}
=\frac{\partial[\theta(\alpha)\psi(q)]}{\partial
t}=\frac{\partial\theta(\alpha)}{\partial
t}\psi(q)+\theta(\alpha)\frac{\partial\psi(q)}{\partial t},
\end{equation},
where the second term in the right side generates the Shrodinger
equation since
\begin{equation}\label{U25}
\theta(\alpha)\frac{\partial\psi(q)}{\partial
t}=\frac{-i\theta(\alpha)}{\hbar}H\psi(q).
\end{equation}

Here $H$ is the Hamiltonian and the first member is added,
similarly to the member that appears in the deformed Loiuville
equation, and  vanishes when $\theta[\alpha(t)]\approx const$. In
particular, this takes place in the low energy limit in QM, when
$\alpha\rightarrow 0$.  It should be noted that the
above theory  is not a time reversal of QM as the
combination $\theta(\alpha)\psi(q)$ breaks down this property in
the deformed Shrodinger equation. Time-reversal is conserved
only in the low energy limit, when a quantum mechanical Shrodinger
equation is valid.
\section{Some Comments and Unitarity in QMFL}
As has been indicated in the previous section,
time reversal is retained in the large-scale limit only.
The same is true for the superposition principle in Quantum Mechanics:
indeed it may be retained in a very narrow interval of cases
for the functions $\psi_{1}(\alpha,q)=\theta(\alpha)\psi_{1}(q)$
è $\psi_{2}(\alpha,q)=\theta(\alpha)\psi_{2}(q)$ with the same value
$\theta(\alpha)$. However, as for all $\theta(\alpha)$, their limit is
$\lim\limits_{\alpha\rightarrow 0}|\theta(\alpha)|^{2}=1$ or equivalently
$\lim\limits_{\alpha\rightarrow
0}|\theta(\alpha)|=1$, in going to the low-energy
limit each wave function $\psi(q)$ is simply multiplied by the phase
factor $\theta(0)$. As a result we have
Hilbert Space wave functions in QM.
Comparison of both pictures (Neumann's and Shrodinger's),
is indicative of the fact that the unitarity means
the retention of the probabilities
$\omega_{i}(\alpha)$ or retention of the squared modulus
(and hence the modulus) for the function $\theta(\alpha)$:
$|\theta(\alpha)|^{2}$,($|\theta(\alpha)|$).That is
\\
\\
$$\frac{d\omega_{i}[\alpha(t)]}{dt}=0$$ or
$$\frac{d|\theta[\alpha(t)]|}{dt}=0.$$
\\
\\
In this way a set of unitary transformations
of QMFL includes a group
$U$ of the unitary transformations for the wave functions
$\psi(q)$ in QM.
\\It is seen that on going from Planck's scale to the
conventional one , i.e. on transition from the early Universe to
the current one, the scale has been rapidly changing in the
process of inflation expansion and the above conditions
failed to be fulfilled:
\begin{equation}\label{U26}
\frac{d\omega_{i}[\alpha(t)]}{dt}\neq 0, {\sloppy}
\frac{d|\theta[\alpha(t)]|}{dt}\neq 0.
\end{equation}
In terms of the density pro-matrices of section 2 this is a
limiting transition from the density pro-matrix in QMFL
$\rho(\alpha)$,$\alpha>0$ , that is a prototype of the pure state
at $\alpha\rightarrow 0$, to the density matrix $\rho(0)=\rho$
representing a pure state in QM. Mathematically this means that a
nontotal probability (below 1) is changed by the total one (equal
to 1). For the wave functions in Schrodinger picture this limiting
transition from QMFL â QM is as follows:
\\
\\
$$\lim\limits_{\alpha\rightarrow 0}\theta(\alpha)\psi(q)=\psi(q)$$
up to the phase factor
\\
\\
It is apparent that the above transition from QMFL to QM is
not a unitary process, as indicated in
\cite{r3},\cite{r4},\cite{r8}-\cite{r10}.
However, the unitarity may be recovered when we consider
in a sense a reverse process:
absorption of the matter by a black hole and its transition to singularity
that conforms to the reverse and nonunitary transition from QM to QMFL.
Thus, nonunitary transitions occur in this picture twice:
\\
\\
$$I.(QMFL,OS,\alpha\approx 1/4)\stackrel{Big\enskip
Bang}{\longrightarrow}(QM,\alpha\approx 0)$$
\\
\\
$$II.(QM,\alpha\approx 0)\stackrel{absorbing\enskip BH
}{\longrightarrow}(QMFL,SBH,\alpha\approx 1/4),$$
\\
\\
Here the following abbreviations are used:
OS for the Origin Singularity; BH for a Black
Hole ; SBH for the Singularity in Black Hole.
\\
As a result of these two nonunitary transitions the total
unitarity may be recovered:
\\
\\
$$III.(QMFL,OS,\alpha\approx
1/4){\longrightarrow}(QMFL,SBH,\alpha\approx 1/4)$$
\\
\\
In such a manner the total information quantity in the Universe
remains unchanged, i.e. no information loss occurs.
\\ In terms of the deformed Liouville equation \cite{r8}-\cite{r10}
we arrive to the expression with the same right-hand parts for
$t_{initial}\sim t_{Planck}$ and $t_{final}$  (for $\alpha\approx
1/4$).
\begin{eqnarray}\label{U27}
\frac{d\rho[\alpha(t),t]}{dt}=\sum_{i}
\frac{d\omega_{i}[\alpha(t)]}{dt}|i(t)><i(t)|-\nonumber \\
-i[H,\rho(\alpha)]= d[ln\omega(\alpha)]\rho
(\alpha)-i[H,\rho(\alpha)].
\end{eqnarray}
It should be noted that for the closed Universe one can consider Final Singularity (FS) rather than the Singularity of Black Hole (SBH), and
then the right-hand parts of the diagrams
II è III will be changed:
\\
\\
$$IIa.(QM,\alpha\approx 0)\stackrel{Big\enskip Crunch
}{\longrightarrow}(QMFL,FS,\alpha\approx 1/4),$$
\\
\\
$$IIIa.(QMFL,OS,\alpha\approx
1/4){\longrightarrow}(QMFL,FS,\alpha\approx 1/4)$$
\\
\\
At the same time, in this case the general unitarity and information
are still retained, i.e. we again have the "unitary" product of
two "nonunitary" arrows:
\\
\\
$$IV.(QMFL,OS,\alpha\approx 1/4)\stackrel{Big\enskip
Bang}{\longrightarrow}(QM,\alpha\approx 0)\stackrel{Big\enskip
Crunch }{\longrightarrow}(QMFL,FS,\alpha\approx 1/4),$$
\\
\\
Finally, arrow III may appear immediately, i.e. without the
appearance of arrows I and II, when in the Early Universe mini BH
are arising:
\\
\\
$$IIIb.(QMFL,OS,\alpha\approx
1/4){\longrightarrow}(QMFL,mini\enskip BH, SBH,\alpha\approx
1/4)$$
\\
\\
However, here, unlike the previous cases, the unitary transition
occurs immediately, without any additional nonunitary ones and
with retention of the total information.
\\ And in terms of the entropy density matrix introduced in
\cite{r11},
\\
\\
$$S^{\alpha_{2}}_{\alpha_{1}}=-Sp[\rho(\alpha_{2})\ln(\rho(\alpha_{1}))]=
-<\ln(\rho(\alpha_{1}))>_{\alpha_{2}},$$
\\
\\
retention of the information means that for any observer
in the proper measurement scale $x_{2}$(with the deformation parameter
$\alpha_{2}$)the densities of entropy in the vicinity of
the initial and final singularity
($\alpha_{1}\approx 1/4$) are coincident:
\\
$$S(in)=S(out)=S^{\alpha_{2}}_{1/4}.$$

\section {Unitarity, Non-Unitarity and Heisenberg's Algebra Deformation}

The above-mentioned unitary and nonunitary transitions may be
described in terms of Heisenberg's algebra deformation
(deformation of commutators) as well. We use the principal results
and designations from \cite{r5}.In the process the following
assumptions are resultant: 1)The three-dimensional rotation group
is not deformed; the angular momentum ${\bf J}$ satisfies the
undeformed $SU(2)$ commutation relations, and the coordinate and
momenta satisfy the undeformed commutation relations $\left[
J_i,x_j\right] =i\epsilon_{ijk}x_k, \left[ J_i,p_j\right]
=i\epsilon_{ijk}p_k$. 2) The momenta commute between themselves:
$\left[ p_i,p_j\right] =0$, so the translation group is also not
deformed. 3) The $\left[ x,x\right]$ and $\left[ x,p\right]$
commutators depend on the deformation parameter $\kappa$ with
dimensions of mass. In the limit $\kappa\rightarrow \infty$ when
$\kappa$ is much larger than any energy, the canonical commutation
relations are recovered.
\\
For a specific realization of points 1) to 3) the generating GUR
are of the form \cite{r5}: ($\kappa$-deformed Heisenberg algebra)
\begin{eqnarray}
\left[ x_i ,x_j \right] &= & -\frac{\hbar^2}{\kappa^2}\,
i\epsilon_{ijk}J_k\label{xx}\\ \left[ x_i , p_j \right]   &= &
i\hbar\delta_{ij} (1+\frac{E^2}{\kappa^2})^{1/2}\, .\label{xp}
\end{eqnarray}
Here $E^2=p^2+m^2$. Note that in this formalism
the transition from GUR to UR, or equally from
QMFL to QM with $\kappa\rightarrow
\infty$ ,from Planck scale to the conventional one, is nonunitary, exactly
following the transition from density pro-matrix to the density
matrix in previous sections:
\\
$$\rho(\beta\neq 0)\stackrel{\beta\rightarrow
0}{\longrightarrow}\rho$$
\\
Then the first arrow I in the formalism of this section
may be as follows:
\\
$$I^{\prime}.(GUR,OS,\kappa\sim M_{p})\stackrel{Big\enskip
Bang}{\longrightarrow}(UR,\kappa=\infty)$$ or what is the same
$$I^{\prime\prime}.(QMFL,OS,\kappa\sim M_{p})\stackrel{Big\enskip
Bang}{\longrightarrow}(QM,\kappa=\infty),$$
\\
\\
where $M_{p}$ is the Planck mass.
In some works of the last two years Quantum Mechanics for a
black hole has been already considered as a Quantum Mechanics
with GUR \cite{r13}-\cite{r15}. As a consequence, by this approach
the black hole is not completely evaporated but rather some stable
remnants always remain in the process
of its evaporation with a mass $\sim
M_{p}$. In terms of \cite{r5} this means nothing else but a reverse
transition:
$(\kappa=\infty)\rightarrow(\kappa\sim M_{p})$. And for an outside
observer this transition is of the form:
\\
$$II^{\prime}.(UR,\kappa=\infty)\stackrel{absorbing\enskip
BH}{\longrightarrow}(GUR,SBH,\kappa\sim M_{p}),$$ òî åñòü
$$II^{\prime\prime}.(QM,\kappa=\infty)\stackrel{absorbing\enskip
BH}{\longrightarrow}(QMFL,SBH,\kappa\sim M_{p}),$$
\\
\\
So similar to the previous section, two nonunitary mutually
reverse transitions:
a)$I^{\prime},(I^{\prime\prime})$;b)$II^{\prime},(II^{\prime\prime})$
are liable to generate a unitary transition:
\\
\\
$$III^{\prime}.(GUR,OS,\kappa\sim M_{p})\stackrel{Big\enskip
Bang}{\longrightarrow}(UR,\kappa=\infty)\stackrel{absorbing\enskip
BH}{\longrightarrow}(GUR,SBH,\kappa\sim M_{p})$$
\\
or to summerize
\\
$$III^{\prime\prime}.(GUR,OS,\kappa\sim
M_{p})\rightarrow(GUR,SBH,\kappa\sim M_{p})$$
\\
In conclusion of this section it should be noted that
an approach to the Quantum Mechanics at Planck Scale
using the Heisenberg algebra deformation (similar to the approach
based on the density matrix deformation from the
previous section) gives a deeper insight into the possibility of
retaining the unitarity and the total quantity of information in
the Universe, making possible the solution of Hawking's information paradox
\cite{r16}-\cite{r18}.

\section{Conclusion}

Thus, this work outlines that the existence of GUR
and hence the appearance of QMFL not only enable a better
understanding of the information problem in the Universe but also
provides a key to the solution of this problem in a not
inconsistent manner practically in the same way
but irrespective of the approach: density matrix deformation or
Heisenberg algebra deformation.
\\
It should be noted that the question of the relationship between these two
approaches, i.e. transition from one deformation to the other, still remains open. This aspect is to be studied in further investigations of the author.

\end{document}